# Initial stage of the 2D-3D transition of a strained SiGe layer on a pit-patterned Si(001) template

Gang Chen, Herbert Lichtenberger, Guenther Bauer, Wolfgang Jantsch, and Friedrich Schäffler

*Institut für Halbleiter- und Festkörperphysik, Johannes Kepler Universität Linz, Austria*

We investigate the initial stage of the 2D-3D transition of strained Ge layers deposited on pit-patterned Si(001) templates. Within the pits, which assume the shape of inverted, truncated pyramids after optimized growth of a Si buffer layer, the Ge wetting layer develops a complex morphology consisting exclusively of {105} and (001) facets. These results are attributed to a strain-driven step-meandering instability on the facetted side-walls of the pits, and a step-bunching instability at the sharp concave intersections of these facets. Although both instabilities are strain-driven, their coexistence becomes mainly possible by the geometrical restrictions in the pits. It is shown that the morphological transformation of the pit surface into low-energy facets has strong influence on the preferential nucleation of Ge islands at the flat bottom of the pits.

Ge on Si(001) is considered a model system for strain-driven (Stranski-Krastanov) 3D island growth, and many of the basic properties, such as island faceting,[1,2,3,4] and alloying by interdiffusion[5,6] were most thoroughly investigated in this group-IV heterosystem. This widespread interest was to a large extent triggered by potential electronic and optoelecronic applications of Si/SiGe heterostructures, and its compatibility with the mature Si-technology. Most recently, laterally ordered Ge quantum dots were proposed as building blocks for quantum computing,[7] and quantum information storage.[8] Both applications require perfectly ordered Ge dots to allow their separate identification and external addressability.

Self-organized Ge islands are usually randomly arranged when grown on a plain substrate. To obtain laterally ordered Ge/Si quantum dot arrays, various approaches were presented in the literature, such as the growth in $SiO_2$ windows,[9,10] or growth above periodic strain fields.[11,12] Recently, almost perfectly ordered arrays of Ge islands with high size homogeneity were achieved by selective nucleation on periodically modulated, one-dimensional[13,14] (1D) and two-dimensional[15,16,17] (2D) patterned Si(001) substrates. The surface of a 2D-patterned substrate is composed of unit cells of faceted pits,[16] where in the ideal case only a single island is formed at the bottom of each pit. This results in an array of islands that reproduces the pattern with the accuracy of the lithographic technique employed for its definition.[9,10,15,16,17]

In spite of these technical achievements, neither the preferential nucleation of Ge dots within the pits, nor the role of the faceted pit sidewalls during the initial states of heterolayer deposition are well understood. Several recent publications indicate that strain relaxation and the nucleation of morphological corrugations and/or islands on vicinal surfaces is considerably more complicated than on flat surfaces. Extending earlier work of Teichert *et al*.[18], Lichtenberger and coworkers [19] showed that, upon overgrowth with compressively strained SiGe, vicinal Si(001) surfaces with local

**2**

inclination angles around 8° in [110] direction become completely {105} faceted in the inclined areas via a step meandering instability. Because of geometrical restrictions, the miscut areas disintegrate into [110] oriented prisms with triangular cross section that are bounded by two adjacent {105} facets. This result appears to be closely related to reports by Ronda et al.,[20] who found the formation of a wire-like, micro-{105}-faceted morphology along the [110] miscut direction of Si(001) substrates that are as a whole misoriented by about 10°. In another recent set of experiments, Watanabe et al. deposited SiGe films on Si(001) substrates that were textured by laser ablation with micron-sized, rotational-symmetric dimples.[21,22] These provide a continuous distribution of miscut angles from 0° to ≈15°, and the complete range of miscut azimuths. Under optimized growth conditions[22] they found for miscuts between ≈5° and ≈12° a radial, ripple-shaped morphology oriented along the four equivalent <110> directions, and island formation outside this miscut range.[21] Although not explicitly investigated by Watanabe et al., the symmetry of their ripples suggest that they are of the same origin as the ones observed in Ref. 20, and that they are therefore most likely also made up of {105} micro-facets.

In all three cases, the morphology of the compressively strained $Si_{1-x}Ge_x$ layer is driven by the same mechanisms that lead on flat Si(001) surfaces to the well-known formation of {105} faceted pyramids or hut clusters.[23] These are elastic strain relaxation via 3D growth and simultaneous minimization of the surface energy by the formation of the low-energy[24] {105} facets. In contrast to a flat Si(001) surface, however, the symmetry of the island becomes more and more distorted as the local miscut increases, and reaches the form of 1D wires, if the miscut approaches the inclination angle of the <551> intersection line between two adjacent {105} facets.[19]

Thus, on a macroscopic (micron-scale) level the morphological evolution and symmetry of strained SiGe layers on (locally) vicinal Si(001) surfaces is at least

**3**

qualitatively reasonably well understood. In this paper, we investigate on a microscopic scale the initial process of the 2D-3D transition of strained SiGe layers on pit-patterned Si(001) templates with typical dimensions in the 100 nm range, and their implications for island nucleation in the pits.

The samples were grown by solid source molecular beam epitaxy (MBE) in a Riber SIVA 45 reactor.[13] Pit-patterned templates on Si(001) substrates were obtained by electron beam lithography and reactive ion etching (RIE) in $SF_6$. The pits form a regular two-dimensional grid ordered along two orthogonal <110> directions with a periodicity of between 260 and 380 nm. After etching, the samples were chemically cleaned to remove RIE deposits and residues from the e-beam resist. Immediately before introduction into the MBE reactor the samples were treated in diluted HF to create a hydrogen terminated surface that stabilizes the nanostructures against transient enhanced diffusion[25] during the following *in situ* thermal treatment at ~900$^o$C for 5 minutes. Since we wanted to characterize the very early stages of ordered Stranski-Krastanov growth on prepatterned substrates, we employed growth conditions that were optimized in our group for this purpose over the last few years:[13,15,16,17] MBE growth commenced with a 100 nm thick Si buffer layer that was grown at a rate of 0.5 Å/s while ramping the substrate temperature from 450$^o$C to 520$^o$C. This procedure eliminates surface roughness and damage induced by RIE, but, most importantly, modifies the morphology and improves the homogeneity of the pattern, as discussed below. Subsequently, 0 (reference for the effect of the Si buffer alone), 2.6, 4, and 5 monolayers (ML) of Ge were deposited at 620$^o$C and at a fixed rate of 0.03 Å/s on samples A, B, C, and D, respectively. Since this rate is at the lower limit of our flux controller, we further enhanced the migration of the deposited Ge atoms or dimers by growth interruption for 10 s after each deposited ML. Afterwards, the substrate temperature was quickly decreased, and the surface morphology was

**4**

characterized *ex-situ* with a Digital Instruments atomic force microscope (AFM) in the tapping mode.

After RIE both the edge and the bottom of the pits were of round shape, and the depth was around 100 nm.[15] The Si buffer layer has dramatic influence on this geometry: Under the optimized growth conditions employed here it converts the etch pits into less than 15 nm deep inverted pyramids with a truncated apex and sharp concave intersections (SCI) between neighboring sidewalls. This is shown in Figure 1 (a) for four adjacent pits on reference sample A. The sidewalls were found to be {11n} planes, which are inclined by 7±2° relative to the (001) surface [line scan in Figure 1 (b)]. This shape transformation through homoepitaxial growth is a robust feature, which is reproducibly observed at not too high growth temperatures on non-overlapping etch pits with a depth comparable to the thickness of the buffer layer.[16,26] Nevertheless, it is not yet clear what stabilizes the sharp concave intersections against capillary forces.[27] Also, the mentioned variation of the sidewall inclination, which is related to the size fluctuations of the etched pits, makes it rather unlikely that these {11n} sidewalls correspond to known low-energy facets of macroscopic Si crystals. Instead, the appearance and stability of the concave intersections could be related to the concave structures observed during kinetic, homoepitaxial step bunching on vicinal Si(001) miscut along [100], which was tentatively associated with the particular arrangement of the double atomic height steps near a concave intersections of two vicinal Si(001) surfaces.[28]

Deposition of 2.6 ML of Ge on the well-defined Si pits leads to a complex, but highly symmetric pattern of surface corrugations: Figure 2 (a) and (b) show an AFM image and its Laplacian convolution from a typical array of pits on sample B, and Figure 2 (c) shows a line scan across the sidewalls of a pit array along [1$\bar{1}$0]. Labeled details of one of these pits, and a cross sectional profile across the [100] diagonal of

**5**

the pit, are depicted in Fig. 3. At this low Ge coverage two main effects can be distinguished. (i) The {11n} sidewalls of the pits disintegrate into a set of ridges that are oriented along the respective <110> direction of inclination [labeled "R" in Figure 3 (b)]. (ii) The SCIs between neighboring sidewalls are no longer stable. They become filled in by the energetically favorable {105} facets, the azimuths of which are rotated by 45° with respect to the original {11n} sidewalls of the pits. However, we do not observe a single {105} facet in each corner of the pits, but a flight of stairs consisting of short terraces [labeled "T" in Figure 3 (b)] that are connected by {105} slopes, as can be seen in the [100] line scan in Figure 3 (c). The latter also shows that at this low Ge coverage the (001) oriented bottom of the pit is basically conserved by a conformal Ge wetting layer with some indications of roughening.

If the amount of Ge is increased to 4 ML (sample C), both types of corrugations become more pronounced, and a single, {105} faceted Ge pyramid with a height of 6±0.5 nm and a base width of 60±5 nm nucleates at the bottom of the pits (Figure 4). The amplitude of the corrugation pattern is now large enough for a quantitative assignment of facets. For this we extracted topographic surface orientation maps, which show the local surface inclination with respect to the (001) surface [Figure 4(b)]. Also, line scans along the [1$\bar{1}$0] and [100] [Figure 4(c)] directions were recorded to identify faceting in the ridge and terrace areas, respectively. The measured slope angle of the side wall ridges is 8±1°, a value consistent with the 8.05° inclination with respect to the (001) surface of the <551> intersection lines between two adjacent {105} facets. We therefore conclude that the ridges have the shape of prisms with triangular cross section that are bounded by adjacent {105} facets. Obviously, we observe the same phenomenon as described in Ref. 19. The short terraces in the corner regions are identified as (001) facets that are separated by {105} facets. They are in registry with the faceted prisms of the

**6**

sidewalls, i.e., the (001) steps develop at the intersections of two prisms from adjacent sidewalls. As a result, the truncated pyramid structure of the Si pits becomes almost completely covered with corrugated {105} facets and a small percentage of interconnecting (001) facets in the corner regions.

Further increase of the amount of deposited Ge to 5 ML (sample D) leads to the expected conversion of the central Ge pyramid into a multi-faceted Ge dome with a base width of 100±5 nm and a height of 17±2 nm (Figure 5). The corrugated morphology of the pit walls and corners becomes somewhat more pronounced but remains otherwise very similar. At even higher Ge coverage secondary and ternary Ge pyramids nucleate at the (001) steps that decorate the pit corners, as has been reported recently.[26]

To test the reproducibility and stability of the observed morphological evolution of pit-patterned Si substrates upon Ge deposition, we studied a part of sample C where faulty lithography caused significant variations in the etching depth after RIE processing. Although 4 ML of Ge were deposited homogeneously on this sample, we found a striking variation in the status of the morphological evolution, but a high degree of reproducibility of the basic features observed on the more ideal pit patterns of samples B and D. This is shown in Fig. 6, where a set of 6 adjacent unit cells is depicted both as a 3D representation of the AFM image and its Laplace-filtered version. This set was chosen, because it represents the whole bandwidth of pit variations found in this faulty area of the sample. The lower three pits are very similar to what was shown in Figure 4, although the size variation of the central Ge pyramid is more pronounced (side width 60±10 nm). The upper three pits are apparently in an earlier state of evolution that resembles the morphology of sample A in Figures 2 and 3, where the central Ge pyramid has not yet formed. The pit in the upper left corner is barely visible, and has a side wall angle of merely 3.4° along [110] (Figure 6 (c), line

**7**

scan P2). The other two pits in this row have side wall inclinations of between 5.5 and 6.5°, which is still smaller than the 8±1° inclination found in the lower pit row (Figure 6 (c) line scan P1), and on samples B and D. Consequently, the aforementioned sidewall decoration with {105} terminated prisms, which is inherently associated with a sidewall slope of 8.05°, is missing, or highly defective, in the upper three pits, but clearly observed in the Laplace filtered images of the lower pit row [Figure 6 (b)]. On the other hand, the staircase-like feature in the corner region of the pits is present in all pits on sample C, independently of their depth.

Based on these observations we constructed an idealized, schematic 3D illustration of the surface morphology in the pits after the nucleation of the pyramidal Ge island. It is depicted in Figure 7 (a) with illumination from the upper right side for best legibility. The 3D impression is most easily achieved, when first concentrating on the central (outward pointing) Ge pyramid, and then directing one's view toward the staircase in the upper right corner. An enlarged and labeled version of the morphological development in this corner area is shown in Figures 7(b)-(d). The sequence of figures illustrates different mechanisms rather than an evolution in time, and is chosen as a guide for the following discussion.

We start with the disintegration of the $(1\bar{1}n)$ sidewall into prisms bounded by the $(105)$ and $(0\bar{1}5)$ facets, and of the adjacent $(\bar{1}\bar{1}n)$ sidewall into $(\bar{1}05)$ and $(0\bar{1}5)$ terminated prisms[19] [Figure 7 (b)]. These two differently oriented prisms contain a common $(0\bar{1}5)$ facet, which is the one that would fill in the SCI between the two sidewalls to minimize the surface energy of the Ge wetting layer. In the case of perfect registry, the common $(\bar{1}05)$ facets of each intersecting prism pair lie in the same physical plane. This results in a train of short V-shaped grooves in the SCI parts of the original pit [Figure 7 (b)]. They are defined by the $(\bar{1}05)$ and $(105)$ facets of

**8**

corresponding prism pairs, which intersect at 157.4°. The experiments show that these are not stable and become filled in with Ge to form truncated V-grooves that expose (001) facets at the bottom [Figure 7(c)]. This behavior strongly resembles the overgrowth of V-grooves in the (unstrained) GaAs/GaAlAs heterosystem, where it has been described by capillary forces and surface energy minimization.[27]

The formation of the (001) faceted stairs in the regions of the SCIs appears as a straightforward consequence of the disintegration of the sidewalls into {105} faceted prisms. It is, however, not *a priori* clear where and how this process starts. The results of the inhomogeneous part of sample C in Figure 6 indicate that surface corrugations start in the regions of the SCI, because the faceted stairs are even present in the shallower pits, where side-wall faceting into {105} terminated prisms is suppressed or highly defective because of the inadequate inclination angle. This argument is corroborated by the Laplace filtered large scale image of the almost ideal sample B [Fig. 2 (b)]: In some of the pits the AFM images show faulty registry of the sidewall prisms near the center of the sidewalls, which indicates that the surface corrugations start independently from all four corners of the pits. This is plausible, because both strain and material accumulation in the region of the SCIs suggest that the critical thickness for the transition is reached first in these areas.

The formation of {105} faceted corrugations in the {11n} faceted pits provides both elastic strain relaxation and minimized surface energy. It is still interesting to address the growth kinetics that facilitates such a complex and highly symmetric morphological pattern. The {11n} sidewalls of the pits consist in equilibrium of a dense sequence of double atomic height $D_B$ steps,[29] which split into pairs of closely spaced $S_A$ and $S_B$ step pairs during step-flow growth.[30] The formation of {105} faceted prisms on the sidewalls upon deposition of compressively strained Ge can then be attributed to a step-meandering mechanism,[18,19] which is driven by

**9**

elastic strain relaxation and the formation of the low-energy {105} facets. Thus, the initially straight steps meander into a periodic zig-zag pattern that is perfectly correlated from one step to the next, as indicated in Figure 7 (d). Note that this is an idealized view, because the steps on the {105} surfaces are reconstructed and form a smooth plane.[31] The intersection of the prisms in the region of the SCIs leads to a staircase of (001) facets separated by {105} regions, as described above. But, since the (001) regions are per definition step-free, the step-meandering instability of the sidewalls turns into a step-bunching instability in the corner regions when seen along the [010] direction [Figure 7(d)]. In both cases, the number of steps (but not their length) is conserved, i.e. no nucleation barrier has to be surmounted.

The coexistence of strain-driven step bunching and step meandering is a rather unusual constellation that has to our knowledge not yet been discussed in the Si/SiGe heterosystem. Moreover, the theoretically predicted[32] existence of strain-induced step bunching in this material system has been controversially discussed for years.[18,19,33,34] Indeed, on a slightly vicinal (001) substrate, where Ge islands can readily form, the step bunching mechanism, which provides strain relaxation in one dimension only, has to compete with the much more efficient 2D strain relaxation via islands. On the pit patterned substrates the geometrical situation is quite different: On the {11n} sidewalls of the pits the {105} faceted Ge islands degenerate into 1D prisms for purely geometrical reasons. In contrast to extended step bunches on a slightly vicinal substrate, which are oriented perpendicular to the miscut direction, the prisms, which are oriented in miscut direction, relax strain in both directions: Across the prism because of the finite, triangular cross section, and along the prism because of the dense train of steps associated with the prism's 8.05° inclination with respect to the (001) reference plane. The existence of step bunching in the corner regions is related to the formation of prisms on the sidewalls, as we have seen, but the upper pit row in

**10**

Figure 6 shows that this can also occur independently, if the sidewall angles are to shallow. In contrast to 1D step bunching on a slightly vicinal substrate, however, the confined geometry in the pit's corners always induces a finite lateral extension of the step bunches, and thus again the feasibility of strain relaxation both parallel and perpendicular to the bunches. Hence, it appears that the predicted strain relaxation by step bunching can indeed be observed in the Si/SiGe heterosystem, if geometrical restrictions, as in the present case, limit the lateral extension of the step bunches.

Figures. 2, 3 and 6 clearly show that the critical thickness for the activation of the morphological corrugations is smaller than that for the nucleation of the central, also {105}bounded, Ge pyramid. But once this second critical thickness is exceeded, Ge islands nucleate only in the pits, not on the (001) land areas between the pits, provided the spacing of the pits does not exceed the mean-free path of the Ge ad-particles.[15,17] It has been argued that the inclined sidewalls of the pit funnel Ge ad-particles preferentially down to the bottom of the pits, where material becomes accumulated. In a thermodynamic model, this effect is described by capillarity,[27] which is driven by the minimum of the electrochemical potential at the bottom of the pit. This gradient in the electrochemical potential occurs, because at a concave intersection of sidewall facets the number of kink positions is increased, and thus the binding energies of adatoms at this position. The growth kinetics therefore leads to preferential growth at the bottom of the pit,[13] provided ad-particle diffusion on the sidewalls is sufficiently high. On an atomic scale there are also strong indications of a migration asymmetry for the ad-particles: It is now well established that vicinal Si(001) surfaces are unstable against kinetic step bunching,[30] which is always associated with an overall downward current of the ad-particles.[35] However, kinetic Monte Carlo simulations showed that this asymmetry is rather small,[30] and therefore most of the Ge ad-particles are expected to become incorporated into the dense train

**11**

of steps at the {11n} sidewalls and the SCIs. This situation should drastically change, once the sidewalls are completely terminated by {105} facets. Recent *ab initio* calculations of the Ge {105} surface revealed that the potential fluctuations on this reconstructed surface are very smooth, and that therefore diffusion of Ge ad-particles is fast and almost isotropic.[36] As a consequence, one can expect significantly enhanced material transport to the minimum of the electrochemical potential at the bottom of the pits, even though the lateral confinement at this site makes elastic strain relaxation less efficient than elsewhere. Once nucleation has occurred, the Ge island collects most of the ad-atoms within the unit cell of the pattern, and thus grows very rapidly until elastic strain relaxation becomes so inefficient that secondary dots can nucleate.[26]

This line of arguments is strongly supported by the morphology of the inhomogeneous part of sample C in Figure 6: After the deposition of 4 ML of Ge a central Ge island has only nucleated in those pits that have their sidewalls decorated with prism-shaped {105} facets. In the shallower pits [profile P2 in Figure 6 (c)], where this kind of morphological evolution is geometrically suppressed or highly defective, no Ge islands are observed at this coverage. It appears that the {105} terminated side-wall prisms are more efficient pathways for transporting Ge down to the bottom of the pits than the step-bunching regions in the corners, which are present in all pits. This means that the (001) terraces in these otherwise also {105} terminated step bunching regions effectively hamper adatom diffusion toward the bottom of the pits. This has important consequences for an efficient fabrication of ordered Ge islands with a high degree of size homogeneity: The pit morphology after Si buffer layer growth needs to be highly uniform, as has, for example, been demonstrated with samples B and D. But even in this ideal case it is difficult to control the size of small Ge islands in the pits, because most of the Ge adatoms within one unit cell of the pit

**12**

pattern become incorporated in the Ge island, once nucleation has occurred. To improve the size variations of small islands, the size of the unit cells, and thus the number of Ge ad-particles that contribute to the Ge island, has to be reduced further.

Finally, our results shine also new light on the experiments performed on textured substrates with a micron-sized, rotational symmetric dimple pattern.[21,22] Despite the differences in size as compared to our pyramidal pits, the symmetry of the ripple morphologies that develop in either case upon overgrowth with Ge or SiGe is strikingly similar. In both cases <110> oriented ridges along the local miscut direction cover the inclined sidewalls of the pits. These ripples are connected by concentric, <100> oriented corrugations, which, although more pronounced in our experiments, can as well be seen on the laser patterned substrates, e.g. in Figure 1 (f) of Ref. 22. This similarity provides strong evidence that the basic mechanisms are the same, even though the related experiments of Ronda et al.[20] suggest that the ripple structure in the laser textured pits consists of {105} micro-facets rather than of extended prisms. Nevertheless, it remains an open question, why the authors of Ref. 22 could quite successfully model the ripple evolution with the Asaro, Tiller and Grinfeld (ATG) model, which is generally considered as unsuited for faceted surfaces.

In summary, we have studied the initial stages of Ge island nucleation on Si(001) substrates that were patterned with pits in the shape of truncated inverted pyramids. We find that the Ge wetting layer develops a complex and highly symmetric ripple morphology in the pit areas before the onset of Ge island nucleation. Pattern formation is driven by elastic strain relaxation and surface energy minimization, but geometrical restrictions do not allow the nucleation of Ge islands on the inclined sidewalls of the pits. Instead, a prism-shaped ripple morphology bounded by {105} facets develops without nucleation barrier by step-meandering. As a direct geometrical consequence, the sharp concave intersections of the pit corners

**13**

become decorated with a staircase of (001) and {105} facets that can be viewed as step-bunching areas. Conversion of the pit sidewalls into {105} facets strongly enhances surface diffusion in the pit area, which then triggers preferential nucleation of Ge islands at the bottom of the pits.

Valuable discussions with Z. Zhong are gratefully acknowledged. We thank H. Hoppe and G.Springholz for help with the AFM measurements. This work was financially supported by GMe (Vienna), FWF (Vienna) via projects 14684, 16631, 16223, and SFB 025, as well as by INTAS via Project 03-51-5015.

**Figure Captions**

Figure 1

(a) AFM image of a pit array prepared by e-beam lithography and RIE, after overgrowth with 100 nm of Si (sample A). (b) Line scan through two pits along the dashed line in (a). Sidewall facets are inclined by 8±2° with respect to the (001) surface of the substrate.

**Figure 2**

(a) AFM image of sample B with 2.6 ML of Ge deposited on an array of pits prepared in the same way as those shown in Fig. 1. (b) Laplacian convolution of the AFM image highlights the ripple morphology and symmetry. (c) Line scan along the [1$\bar{1}$0] direction [dashed line in (a)] showing a side wall inclination of 8±1° with respect to the (001) plane.

**Figure 3**

Details of a single pit from Fig. 2. (a): 3D AFM image and (b): Laplace convolution. The sidewalls have become decorated with [110] oriented ripples (labeled "R"), and the SCIs have disintegrated into faceted flights of stairs, the terrace parts of which are labeled "T" in Fig. 3 (b). (c) Line scan across the [100] diagonal of the pit [dashed line in (a)], showing that the SCIs have been predominantly filled in by a {105} faceted Ge wetting layer.

**Figure 4**

(a) AFM image of sample C with 4 ML of Ge. A {105} faceted Ge pyramid has nucleated at the bottom of the pit in addition to the morphological changes seen in Figs. 2 and 3. (b) Topographic Surface Orientation Map (SOM) of the pit. (c) Line profiles of the pit along the [100] and [110] directions [dashed lines P1 and P2 in (a), respectively], showing the disintegration of the SCIs into {105} facets and (001) terraces. Arrows mark the position of the Ge pyramid.

**Figure 5**

AFM image (a) and its Laplace convolution (b) of sample D with 5 ML of Ge. (c): Line scan in [100] direction along the dashed line in (a). The central Ge islands have substantially increased in size (base width ≈100 nm), and are now Ge domes terminated by {113}, {15 3 23} and {105} facets, which leads to the more rounded appearance in the Laplacian convolution. The morphological corrugations of the pit walls are more pronounced, but their symmetry remains essentially unchanged.

**Figure 6**

Array of six pits from a part of sample C (4 ML of Ge) with faulty lithography that led to a spread in the depth of the etch pits. (a): 3D AFM image; (b): Laplace transformation; (c) line scans along [110] through the lower (P1) and upper (P2) pit row. Only the lower pit row has developed the Ge pyramids (arrows in profile P1) and the full symmetry of the morphological features both on the pit walls and in the corners. The upper pits are shallower and show merely the stair case-like surface corrugations in the pits' corners.

**Figure 7**

(a) Schematic 3D representation of the pit structure after the surface is converted into {105} and (001) facets. The central pyramid is outward pointing, and the 3D effect is best seen in the upper right corner. (b) Initial stage of facet formation in the upper right corner of the pit. The sidewall facets are converted into {105} terminated,

**18**

prisms that intersect in the corner regions of the original pit, forming trains of short, downward pointing V-grooves. (c) The V-grooves are not stable against capillary forces and become truncated. This leads to a flight of steps defined by the (001) truncation facets. (d) Same as (c) with lines schematically indicating the step distribution on the various facets. Faceting of the sidewalls results from step meandering, whereas the appearance of the (001) facets can be considered as step bunching, when seen along the [010] direction.

**19**

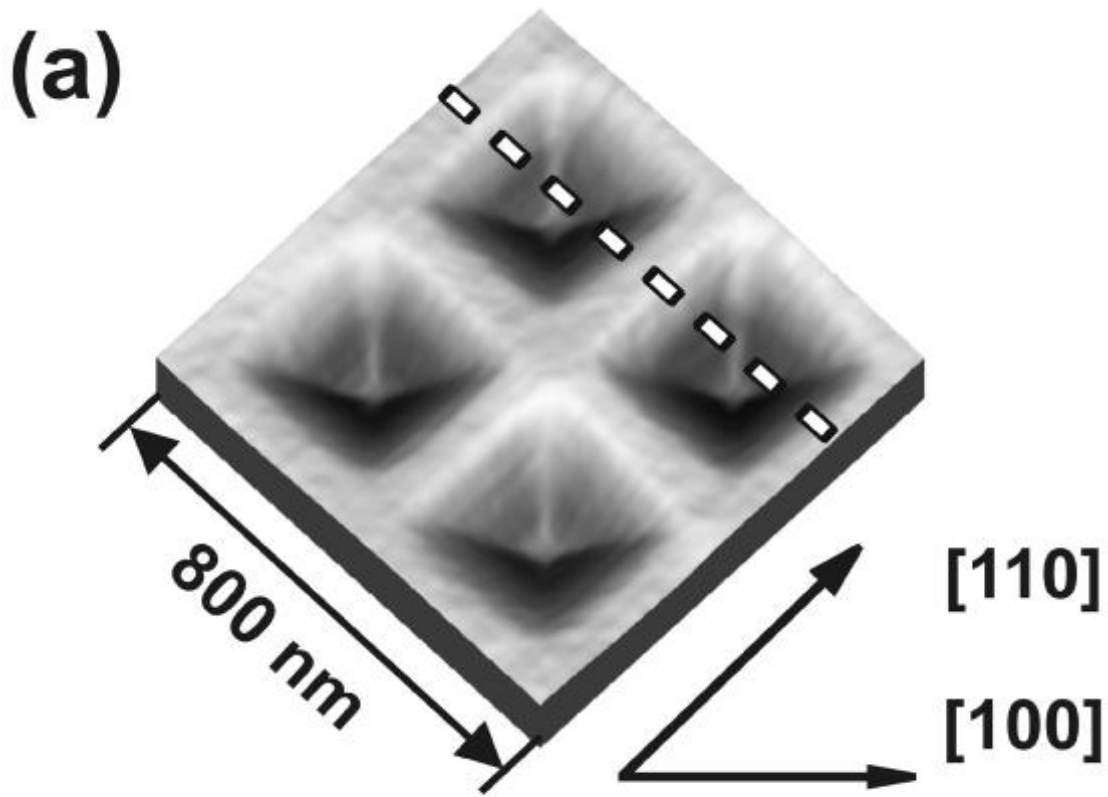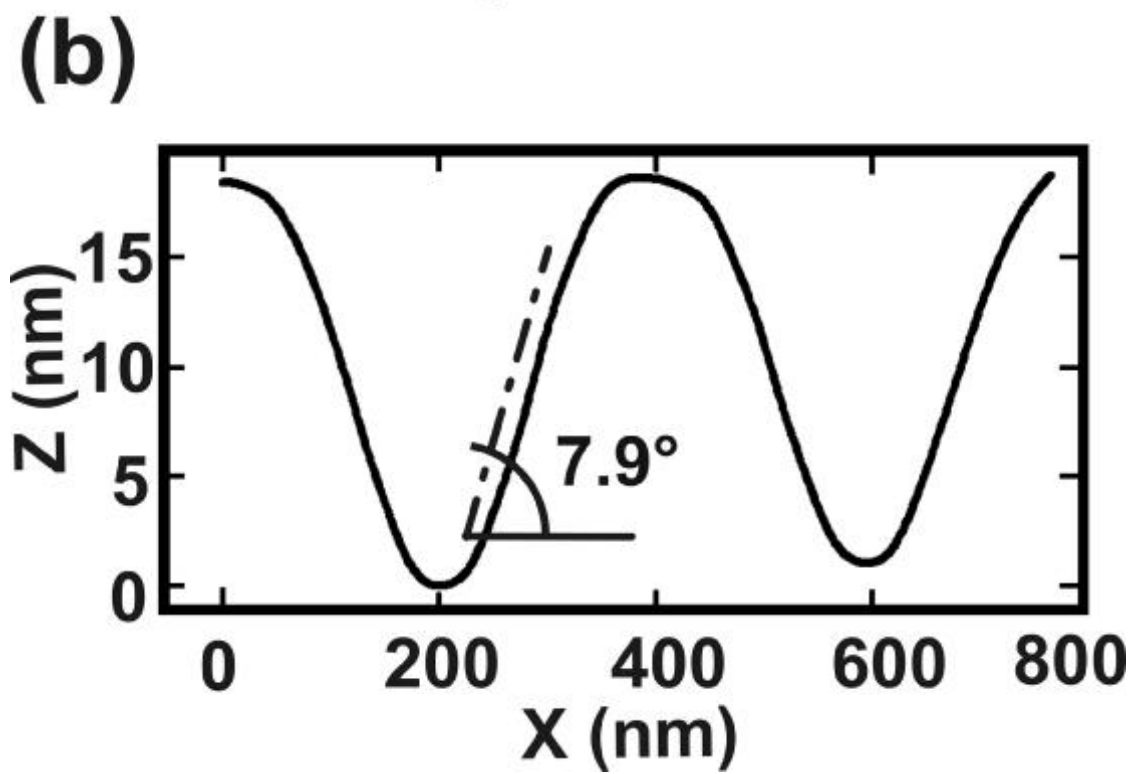

**Chen et al, Figure 1**

**20**

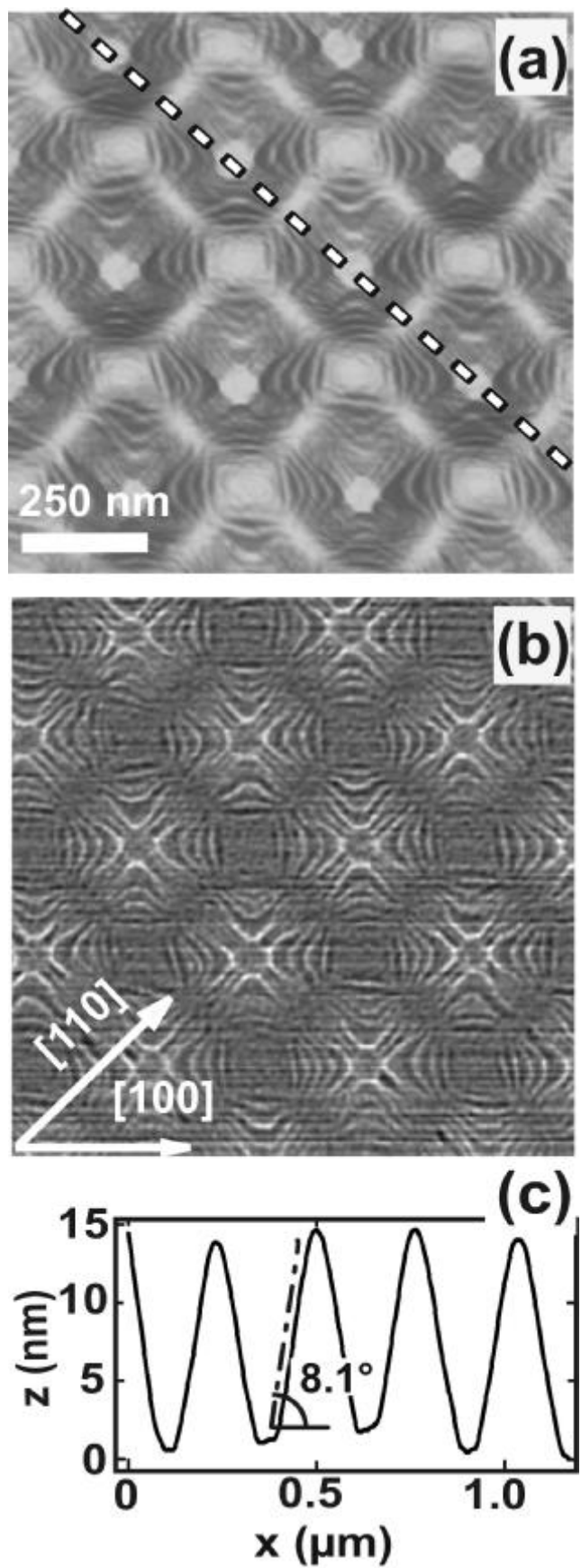

**Chen et al., Figure 2**

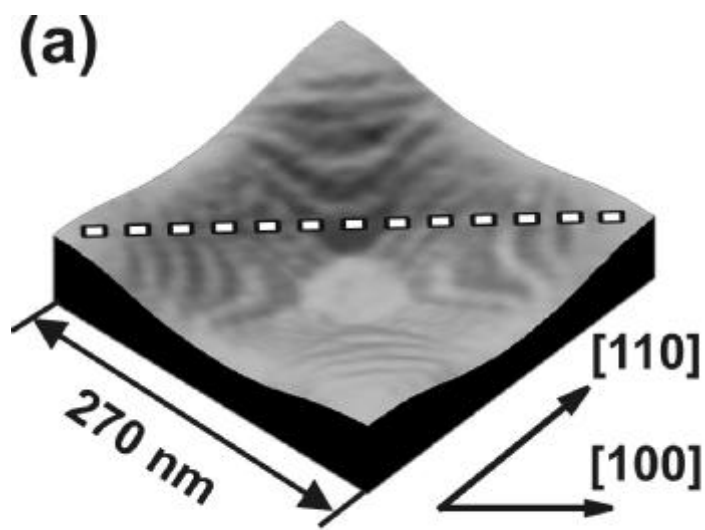

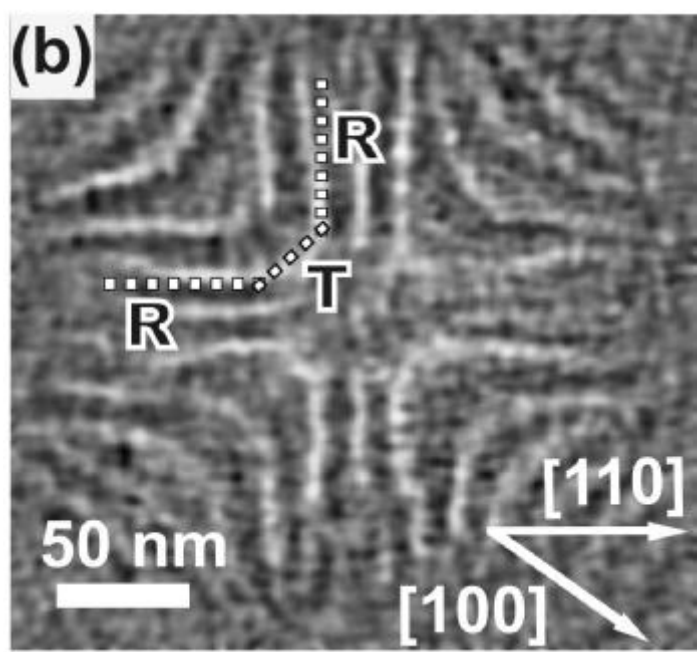

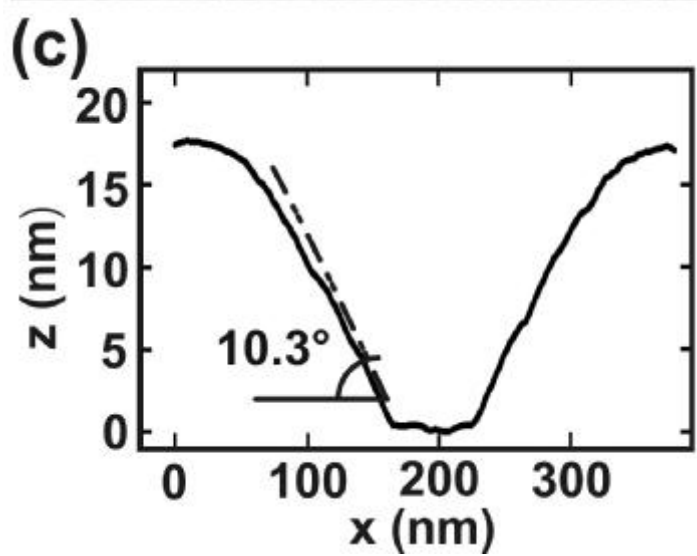

**Chen et al., Figure 3**

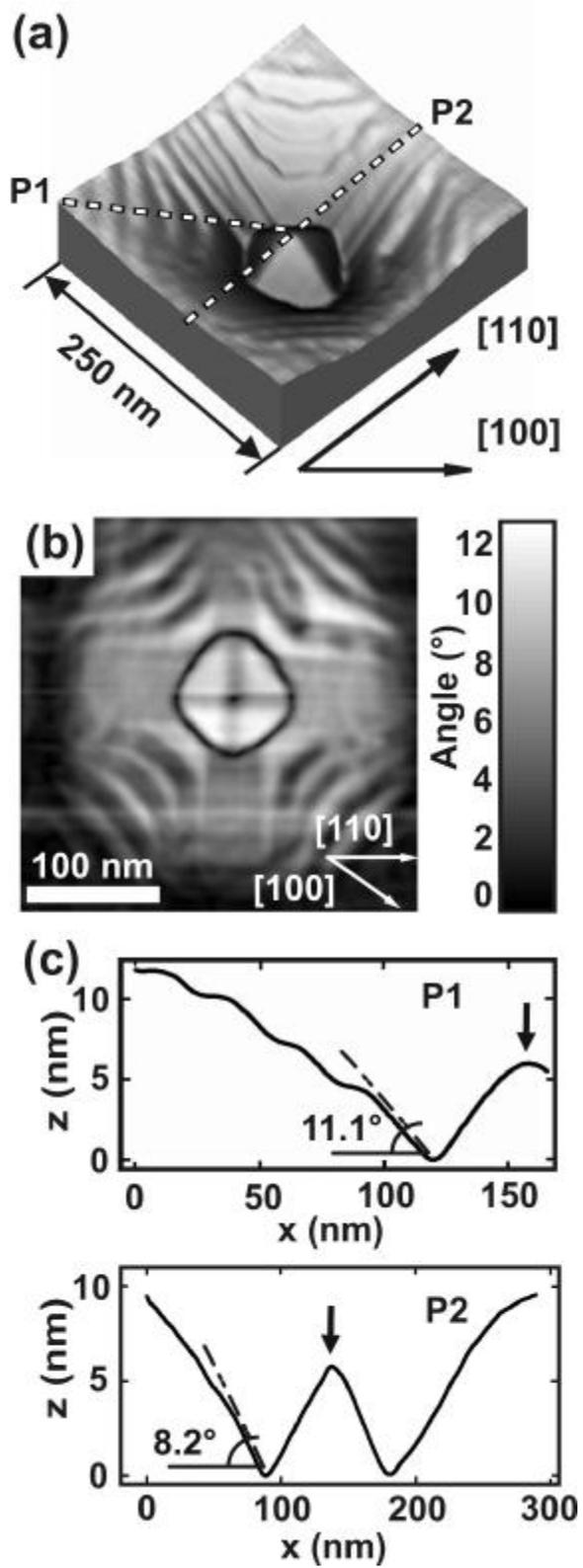

**Chen et al., Figure 4**

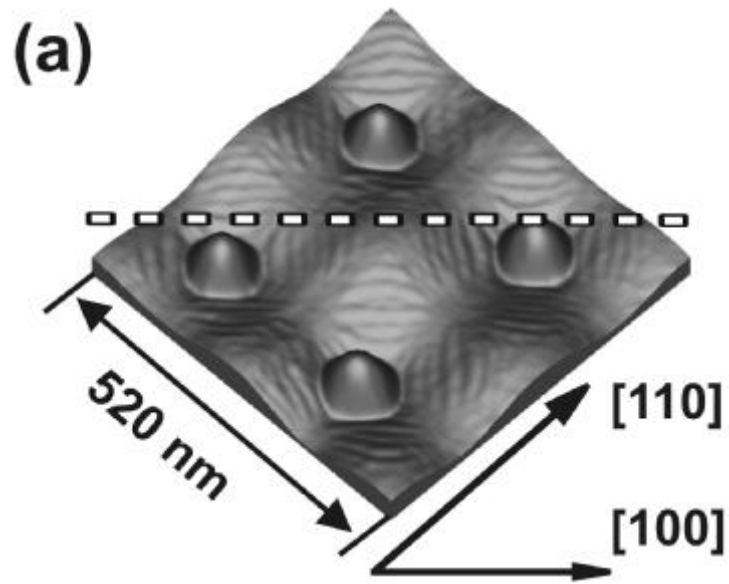

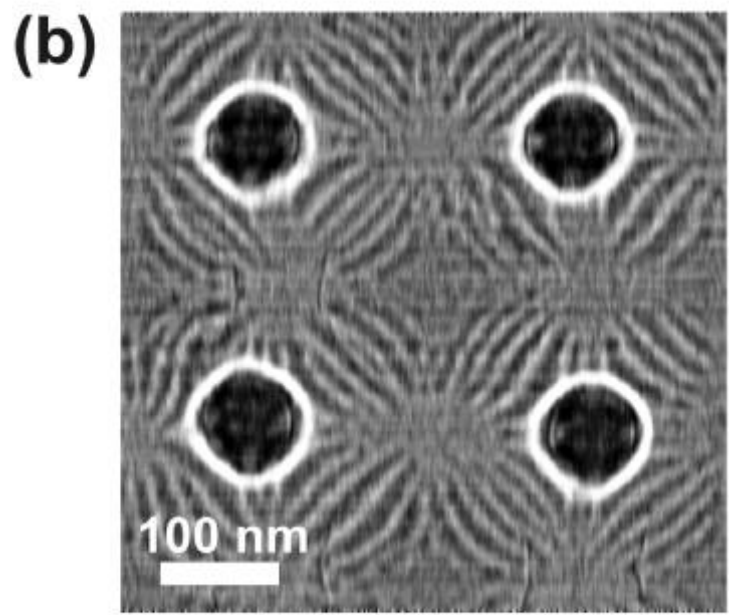

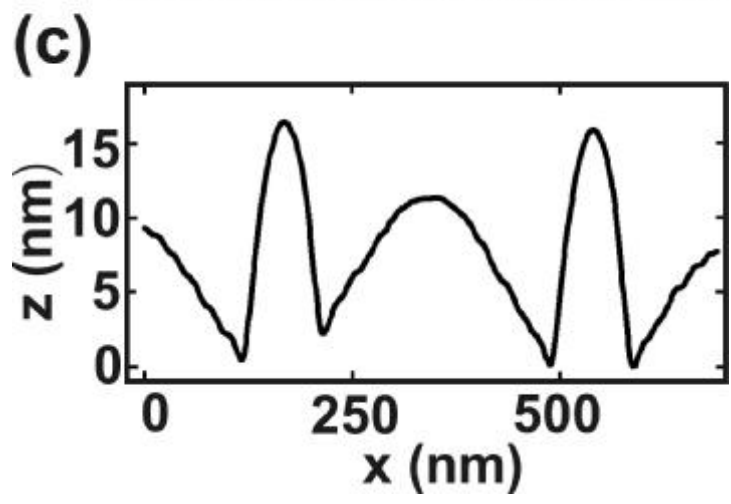

**Chen et al., Figure 5**

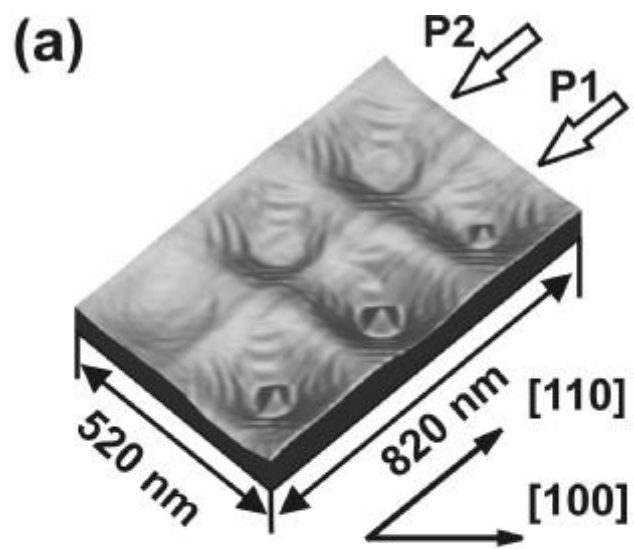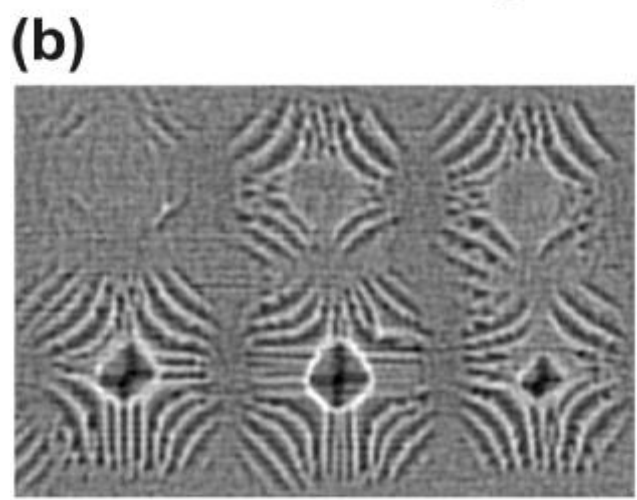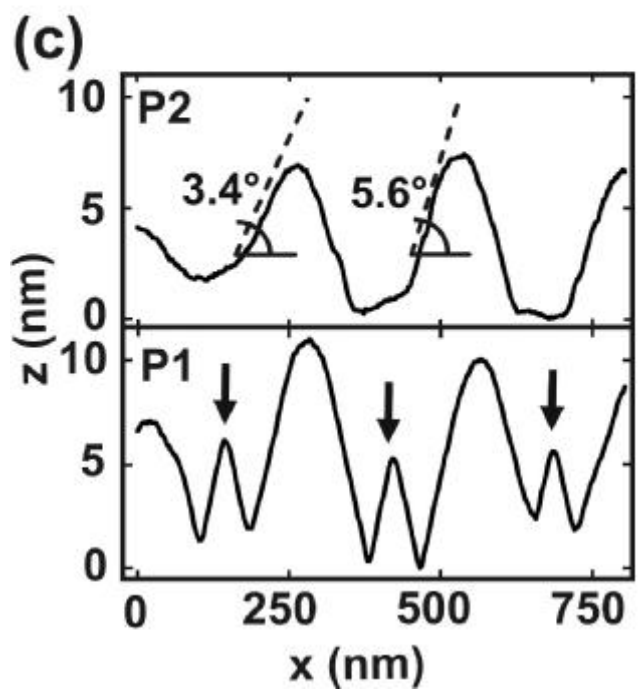

**Chen et al., Figure 6**

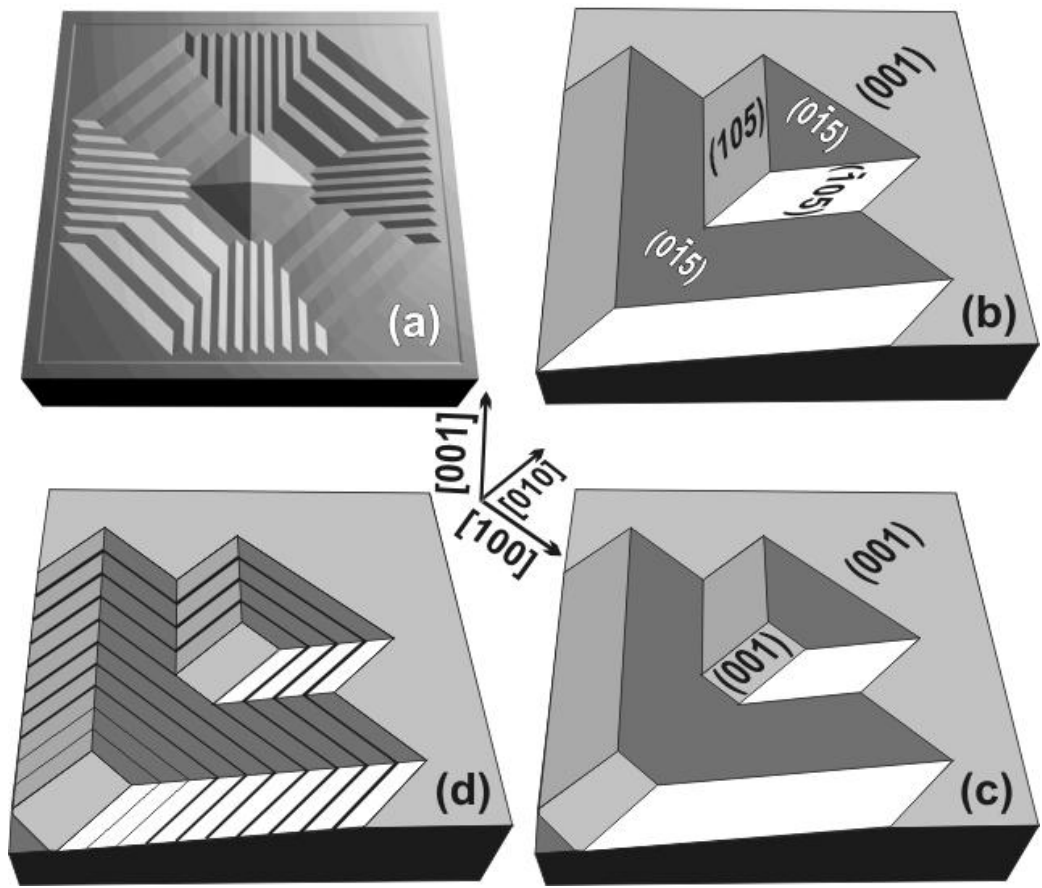

**Chen et al., Figure 7**